\documentclass[twocolumn,prb,amsmath,amssymb,showpacs]{revtex4}
\usepackage{graphicx}
\pdfoutput=1

\begin{document}
\newcommand{\al}{\alpha}
\newcommand{\be}{\beta}
\newcommand{\de}{\delta}
\newcommand{\D}{\Delta}
\newcommand{\e}{\epsilon}
\newcommand{\s}{\sigma}
\newcommand{\del}{\nabla}
\newcommand{\p}{\partial}
\newcommand{\delt}{\partial_t}
\newcommand{\delp}{\nabla_p}
\newcommand{\delq}{\nabla_q}
\newcommand{\bs}[1]{\boldsymbol{#1}}
\newcommand{\mbf}[1]{\mathbf{#1}}
\newcommand{\mcal}[1]{\mathcal{#1}}
\newcommand{\nn}{\nonumber\\}
\newcommand{\up}{\uparrow}
\newcommand{\bea}{\begin{align}}
\newcommand{\eea}{\end{align}}
\newcommand{\ben}{\begin{equation}}
\newcommand{\een}{\end{equation}}
\newcommand{\ban}{\begin{align}}
\newcommand{\ean}{\end{align}}
\newcommand{\bed}{\begin{displaymath} }
\newcommand{\eed}{\end{displaymath}}

\title{Hydrodynamic theory of coupled current and magnetization dynamics in spin-textured ferromagnets}
\author{Clement H. Wong}
\author{Yaroslav Tserkovnyak}
\affiliation{Department of Physics and Astronomy, University of California, Los Angeles, California 90095, USA}

\begin{abstract}
We develop the hydrodynamic theory of collinear spin currents coupled to magnetization dynamics in metallic ferromagnets.  The collective spin density couples to the spin current through a U(1) Berry-phase gauge field determined by the local texture and dynamics of the magnetization. We determine phenomenologically the dissipative corrections to the equation of motion for the electronic current, which consist of a dissipative spin-motive force generated by magnetization dynamics and a magnetic texture-dependent resistivity tensor. The reciprocal dissipative, adiabatic spin torque on the magnetic texture follows from the Onsager principle. We investigate the effects of thermal fluctuations and find that electronic dynamics contribute to a nonlocal Gilbert damping tensor in the Landau-Lifshitz-Gilbert equation for the magnetization. Several simple examples, including magnetic vortices, helices, and spirals, are analyzed in detail to demonstrate general principles.
\end{abstract}



\maketitle

\section{Introduction}

The interaction of electrical currents with magnetic spin texture in conducting ferromagnets is presently a subject of active research. Topics of interest include current-driven magnetic dynamics of solitons such as domain walls and magnetic vortices,\cite{bergerJAP84,zhangPRL04,tserkovPRB06md,tataraPRP08} as well as the reciprocal process of voltage generation by magnetic dynamics.\cite{bergerPRB86,volovikJPC87,sternPRL92,barnesPRL07,saslowPRB07,tserkovPRB08mt,tserkovPRB09md,yangPRL09} This line of research has been fueled in part by its potential for practical applications to magnetic memory and data storage devices.\cite{wolfSCI01} Fundamental theoretical interest in the subject dates back at least two decades.\cite{bergerPRB86,haldanePRL86,volovikJPC87} It was recognized early on\cite{volovikJPC87} that in the adiabatic limit for spin dynamics, the conduction electrons interact with the magnetic spin texture via an effective spin-dependent U(1) gauge field that is a local function of the magnetic configuration. This gauge field, on the one hand, gives rise to a Lorentz force due to ``fictitious" electric and magnetic fields and, on the other hand, mediates the so-called spin-transfer torque exerted by the conduction electrons on the collective magnetization. An alternative and equivalent view is to consider this force as the result of the Berry phase\cite{berryPRSLA84} accumulated by an electron as it propagates through the ferromagnet with its spin aligned with the ferromagnetic exchange field.\cite{barnesPRL07,tserkovPRB08mt,nBerry} In the standard phenomenological formalism based on the Landau-Lifshitz-Gilbert (LLG) equation, the low-energy, long-wavelength magnetization dynamics are described by collective spin precession in the effective magnetic field, which is coupled to electrical currents via the spin-transfer torques. In the following, we develop a closed set of nonlinear classical equations governing current-magnetization dynamics, much like classical electrodynamics, with the LLG equation for the spin-texture ``field" in lieu of the Maxwell equations for the electromagnetic field.

This electrodynamic analogy readily explains various interesting magnetoelectric phenomena observed recently in ferromagnetic metals. Adiabatic charge pumping by magnetic dynamics\cite{moriyamaPRL08} can be understood as the generation of electrical currents due to the fictitious electric field.\cite{tserkovPRB08tb} In addition, magnetic textures with nontrivial topology exhibit the so-called topological Hall effect,\cite{brunoPRL04,binzPRB06} in which the fictitious magnetic field causes a classical Hall effect. In contrast to the classical magnetoresistance, the flux of the fictitious magnetic field is a topological invariant of the magnetic texture.\cite{volovikJPC87}

Dissipative processes in current-magnetization dynamics are relatively poorly understood and are of central interest in our theory. Electrical resistivity due to quasi-one-dimensional (1D) domain walls and spin spirals have been calculated microscopically.\cite{viretPRB96,marrowsAP05,WicklesPRB09} More recently, a viscous coupling between current and magnetic dynamics which determines the strength of a dissipative spin torque in the LLG equation as well the reciprocal dissipative spin electromotive force generated by magnetic dynamics, called the ``$\beta$ coefficient,"\cite{zhangPRL04} was also calculated in microscopic approaches.\cite{tserkovPRB06md,kohnoJPSJ06,duinePRB07fk} Generally, such first-principles calculations are technically difficult and restricted to simple models. On the other hand, the number of different forms of the dissipative interactions in the hydrodynamic limit are in general constrained by symmetries and the fundamental principles of thermodynamics, and may readily be determined phenomenologically in a gradient expansion. Furthermore, classical thermal fluctuations may be easily incorporated in the theoretical framework of quasistationary nonequilibrium thermodynamics.

The principal goal of this paper is to develop a (semi-phenomenological) hydrodynamic description of the dissipative processes in electric flows coupled to magnetic spin texture and dynamics. In Ref.~\onlinecite{tserkovPRB09md}, we drew the analogy between the interaction of electric flows with quasistationary magnetization dynamics with the classical theory of magnetohydrodynamics.  In our ``spin magnetohydrodynamics," the spin of the itinerant electrons, whose flows are described hydrodynamically, couples to the local magnetization direction, which constitutes the collective spin-coherent degree of freedom of the electronic fluid. In particular, the dissipative $\beta$ coupling between the collective spin dynamics and the itinerant electrons is loosely akin to the Landau damping, capturing certain kinematic equilibration of the relative motion between spin-texture dynamics and electronic flows. In our previous paper,\cite{tserkovPRB09md} we considered a special case of incompressible flows in a 1D ring to demonstrate the essential physics.  In this paper, we establish a general coarse-grained hydrodynamic description of the interaction between the electric flows and textured magnetization in three dimensions, treating the itinerant electron's degrees of freedom in a two-component fluid model (corresponding to the two spin projections of spin-$1/2$ electrons along the local collective magnetic order). Our phenomenology encompasses all the aforementioned magnetoelectric phenomena.
 
The paper is organized as follows.  In Sec.~\ref{QA}, we use a Lagrangian approach to derive the semiclassical equation of motion for itinerant electrons in the adiabatic approximation for spin dynamics. In Sec.~\ref{SCL}, we derive the basic conservation laws, including the Landau-Lifshitz equation for the magnetization, by coarse-graining the single-particle equation of motion and the Hamiltonian. In Sec.~\ref{DISS}, we phenomenologically construct dissipative couplings, making use of the Onsager reciprocity principle, and calculate the net dissipation power. In particular, we develop an analog of the Navier-Stokes equation for the electronic fluid, focusing on texture-dependent effects, by making a systematic expansion in nonequilibrium current and magnetization consistent with symmetry requirements. In Sec.~\ref{TN}, we include the effects of classical thermal fluctuations by adding Langevin sources to the hydrodynamic equations, and arrive at the central result of this paper: A set of coupled stochastic differential equations for the electronic density, current, and magnetization, and the associated white-noise correlators of thermal noise. In Sec.~\ref{EX}, we apply our results to special examples of rotating and spinning magnetic textures, calculating magnetic texture resistivity and magnetic dynamics-generated currents for a magnetic spiral and a vortex. The paper is summarized in Sec.~\ref{SUM} and some additional technical details, including a microscopic foundation for our semiclassical theory, are presented in the appendices.

\section{Quasiparticle action}
\label{QA}

In a ferromagnet, the magnetization is a symmetry-breaking collective dynamical variable that couples to the itinerant electrons through the exchange interaction. Before developing a general phenomenological framework, we start with a simple microscopic model with Stoner instability, which will guide us to explicitly construct some of the key magnetohydrodynamic ingredients. Within a low-temperature mean-field description of short-ranged electron-electron interactions, the electronic action is given by (see appendix~\ref{mba} for details):
\ben
S=\int dtd^3r\hat{\psi}^\dag \left[ i\hbar\partial_t +{\hbar^2\over2m_e}\del^2-\frac{\phi}{2}+{\D\over2}\mbf{m}\cdot\hat{\bs{\s}}\right]\hat{\psi}\,.
\label{S0}
\een
Here, $\Delta(\mathbf{r},t)$ is the ferromagnetic exchange splitting, $\mbf{m} (\mbf{r},t) $ is the direction of the dynamical order parameter defined by $\hbar\langle\hat{\psi}^\dag\hat{\bs{\s}}\hat{\psi}\rangle/2=\rho_s\mbf{m}$, $\rho_s$ is the local spin density, and $\hat{\psi}(\mbf{r},t)$ is the spinor electron field operator. For the short-range repulsion $U>0$ discussed in appendix~\ref{mba}, $\Delta(\mbf{r},t)=2U\rho_s(\mbf{r},t)/\hbar$ and $\phi(\mbf{r},t)=U\rho(\mbf{r},t)$, where $\rho=\langle\hat{\psi}^\dag\hat{\psi}\rangle$ is the local particle number density. For electrons, the magnetization $\mbf{M}$ is in the opposite direction of the spin density: $\mbf{M}=\gamma\rho_s\mathbf{m}$, where $\gamma<0$ is the gyromagnetic ratio. Close to a local equilibrium, the magnetic order parameter describes a ground state consisting of two spin bands filled up to the spin-dependent Fermi surfaces, with the spin orientation defined by $\mbf{m}$. We will focus on soft magnetic modes well below the Curie temperature, where only the direction of the magnetization and spin density are varied, while the fluctuations of the magnitudes are not significant. The spin density is given by $\rho_s=\hbar(\rho_+-\rho_-)/2$ and particle density by $\rho=\rho_++\rho_-$, where $\rho_\pm$ are the local spin-up/down particle densities along $\mbf{m}$. $\rho_s$ can be essentially constant in the limit of low spin susceptibility.

Starting with a nonrelativistic many-body Hamiltonian, the action (\ref{S0}) is obtained in a spin-rotationally invariant form. However, this symmetry is broken by spin-orbit interactions, whose role we will take into account phenomenologically in the following. When the length scale on which $\mbf{m}(\mbf{r},t )$ varies is much greater than the ferromagnetic coherence length $l_c\sim\hbar v_F/\Delta$, where $v_F$ is the Fermi velocity, the relevant physics is captured by the adiabatic approximation. In this limit, we start by neglecting transitions between the spin bands, treating the electron's spin projection on the magnetization as a good quantum number. (This approximation will be relaxed later, in the presence of microscopic spin-orbit or magnetic disorder.) We then have two effectively distinct species of particles described by a spinor wave function $\hat{\psi}'$, which is defined by $\hat{\psi}=\hat{\mathcal{U}}(\mathcal{R})\hat{\psi}'$. Here, $\hat{\mathcal{U}}(\mathcal{R})$ is an SU(2) matrix corresponding to the local spatial rotation $\mathcal{R} (\mbf{r},t)$ that brings the $z$-axis to point along the magnetization direction: $\mathcal{R} (\mbf{r},t)\mbf{z}=\mbf{m} (\mbf{r},t)$, so that $\hat{\mcal{U}}^\dagger(\hat{\bs{\s}}\cdot\mbf{m})\hat{\mcal{U}}=\hat{\s}_z$. The projected action then becomes:
\begin{align}
S&=\int dt\int d^3r\hat{\psi}^{\prime\dag}\left[\left(i\hbar \partial_t+\hat {a}\right)-{\left(-i\hbar\bs{\del}-\hat{\mbf{a}}\right)^2\over 2m_e}\right.\nn
&\left.\hspace{3cm}-\frac{\phi}{2}+{\D\over2}\hat{\s}_z\right]\hat{\psi}^\prime-\int dtF[\mbf{m}]\,,
\label{action}
\end{align}
where
\ben
F[\mbf{m}]=\frac{A}{2}\int d^3r(\p_i\mbf{m})^2
\label{Fex}
\een
is the spin-texture exchange energy (implicitly summing over the repeated spatial index $i$), which comes from the terms quadratic in the gauge fields that survive the projection. In the mean-field Stoner model, the ferromagnetic exchange stiffness is $A=\hbar^2\rho/4m_e$. To broaden our scope, we will treat it as a phenomenological constant, which, for simplicity, is determined by the mean electron density.\cite{nExchange} The spin-projected ``fictitious'' gauge fields are given by
\begin{align}
a_\sigma(\mbf{r},t)&=i\hbar\langle\s|\hat{\mathcal{U}}^\dag\partial_t\hat{\mathcal{U}}|\s\rangle\,,\nn
\mbf{a}_\sigma(\mbf{r},t)&=i\hbar\langle\s|\hat{\mathcal{U}}^\dag\bs{\del}\hat{\mathcal{U}}|\s\rangle\,.
\label{aa}
\end{align}
Choosing the rotation matrices $\hat{\mcal{U}}(\mbf{m})$ to depend only on the local magnetic configuration, it follows from their definition that spin-$\s$ gauge potentials have the form:
\ben
a_\s=-\delt\mbf{m}\cdot\mbf{a}_\s^{\rm mon}(\mbf{m})\,,\,\,\,a_{\s i}=-\partial_i\mbf{m}\cdot\mbf{a}_\s^{\rm mon}(\mbf{m})\,, 
\label{gauge fields}
\een
where $\mbf{a}_\s^{\rm mon}(\mbf{m})\equiv-i\hbar\langle\s|\hat{\mathcal{U}}^\dag\partial_\mbf{m}\hat{\mathcal{U}}|\s\rangle$.   We show in Appendix~\ref{mgf} the well known result (see, e.g., Ref.~\onlinecite{bazaliyPRB98}) that $\mbf{a}_\s^{\rm mon}$ is the vector potential (in an arbitrary gauge) of a magnetic monopole in the parameter space defined by $\mbf{m}$:
\ben
\p_\mbf{m}\times\mbf{a}_\s^{\rm mon} (\mbf{m})=q_\s \mbf{m}\,,
\een
where $q_\s=\s\hbar/2$ is the monopole charge (which is appropriately quantized).

By noting that the action \eqref{action} is formally identical to charged particles in electromagnetic field, we can immediately write down the following classical single-particle Lagrangian for the interaction between the spin-$\s$ electrons and the collective spin texture:  
\ben
L_\s(\mbf{r},\dot{\mbf{r}},t)={m_e\dot{\mbf{r}}^2\over2}+\dot{\mbf{r}}\cdot \mbf{a}_{\s}(\mbf{r},t)+a_\sigma(\mbf{r},t)\,,
\label{electron lagrangian}
\een
where $\dot{\mbf{r}}$ is the spin-$\s$ electron (wave-packet) velocity. To simplify our discussion, we are omitting here the spin-dependent forces due to the self-consistent fields $\phi(\mbf{r},t)$ and $\Delta(\mbf{r},t)$, which will be easily reinserted at a later stage. See Eq.~(\ref{TK}).

The Euler-Lagrange equation of motion for $\mbf{v}=\dot{\mbf{r}}$  derived from the single-particle Lagrangian \eqref{electron lagrangian}, $(d/dt)(\p L_\s/\p\dot{\mbf{r}})=\p L_\s/\p\mbf{r}$, gives
\ben
m_e\dot{\mbf{v}}=q_\s(\mathbf{e}+\mathbf{v}\times\mathbf{b})\,.
\label{Lorentz Force}
\een
The fictitious electromagnetic fields that determine the Lorentz force are
\begin{align}
q_\s e_i&=\p_ia_\s-\p_ta_{\s i}=q_\s \mbf{m}\cdot(\partial_t\mbf{m}\times\partial_i\mbf{m})\,,\nn
q_\s b_i&=\e^{ijk}\p_ja_{\s k}=q_\s {\e^{ijk}\over2}\mbf{m}\cdot(\partial_k\mbf{m}\times\partial_j\mbf{m})\,.
\label{fictitious fields}
\end{align}
They are conveniently expressed in terms of the tensor field strength
\ben
q_\s f_{\mu\nu}\equiv\partial_{\mu} a_{\s\nu}-\partial_{\nu} a_{\s\mu}=q_\s\mbf{m}\cdot(\p_\nu\mbf{m}\times\p_\mu\mbf{m})
\een
by $e_i=f_{i0}$ and $b_i=\e^{ijk}f_{jk}/2$. $\e^{ijk}$ is the antisymmetric Levi-Civita tensor and we used four-vector notation, defining $\p_\mu=(\p_t,\bs{\del})$ and $a_{\s\mu}=(a_\s,\mbf{a}_\s)$. Here and henceforth the convention is to use Latin indices to denote spatial coordinates and Greek for space-time coordinates. Repeated Latin indices $i,j,k$ are, furthermore, always implicitly summed over.

\section{Symmetries and conservation laws}
\label{SCL}

\subsection{Gauge invariance}

The Lagrangian describing coupled electron transport and collective spin-texture dynamics (disregarding for simplicity the ordinary electromagnetic fields) is
\begin{align}
L&(\mbf{r}_p,\mbf{v}_p;\mbf{m},\partial_\mu\mbf{m})\nn
&=\sum_p\left({m_e\mbf{v}_p^2\over2}+\mbf{v}_p\cdot\mbf{a}_{\s}+a_{\s}\right)-\frac{A}{2}\int d^3r(\p_i\mbf{m})^2\nn
&=\sum_p\left({m_e\mbf{v}_p^2\over2}+v^\mu_p a_{\s\mu}\right)-\frac{A}{2}\int d^3r(\p_i\mbf{m})^2\,.
\label{lagrangian}
\end{align}
$v^\mu_p\equiv(1,\mbf{v}_p)$, $\mbf{v}_p=\dot{\mathbf{r}}$, and $\s$ here is the spin of individual particles labelled by $p$. The resulting equations of motion satisfy certain basic conservation laws, due to spin-dependent gauge freedom, space-time homogeneity, and spin isotropicity.

First, let us establish gauge invariance due to an ambiguity in the choice of the spinor rotations $\hat{\mcal{U}}(\mbf{r},t)\to\hat{\mathcal{U}}\hat{\mathcal{U}}^\prime$. Our formulation should be invariant under arbitrary diagonal transformations $\hat{\mcal{U}}^\prime=e^{-if}$ and $\hat{\mcal{U}}^\prime=e^{-ig\hat{\s}_z/2}$ on the rotated fermionic field $\hat{\psi}^\prime$, corresponding to gauge transformations of the spin-projected theory:
\ben
\delta a_{\s\mu}=\hbar\p_\mu f\,\,\,\,{\rm and}\,\,\,\delta a_{\s\mu}=\s\hbar\p_\mu g/2\,,
\label{gauge}
\een
respectively.  The change in the Lagrangian density is given by
\ben
\delta\mcal{L}=j^\mu\p_\mu f\,\,\,{\rm and}\,\,\,\delta\mcal{L}=j_s^\mu\p_\mu g\,,
\een
respectively, where $j=j_++j_-$ and $j_s=\hbar(j_+-j_-)/2$ are the corresponding charge and spin gauge currents.  The action $S=\int dt d^3r\mcal{L}$ is gauge invariant, up to surface terms that do not affect the equations of motion, provided that the four-divergence of the currents vanish, which is the conservation of particle number and spin density:
 \begin{align}
 \dot{\rho}+\bs{\del}\cdot\mbf{j}=0\,,\,\,\,\dot{\rho}_s+\bs{\del}\cdot\mbf{j}_{s}=0\,.
 \label{CE}
 \end{align}
(The second of these conservation laws will be relaxed later.) Here, the number and spin densities along with the associated flux densities are
\begin{align}
\rho&=\sum_p  n_p\equiv\rho_++\rho_- \,,\nn
\mathbf{j}&=\sum_p n_p \mbf{v}_p\equiv\rho\mbf{v}\,,
\end{align}
and
\begin{align}
\rho_s&=\sum_p q_\s n_p\equiv{\hbar\over2}(\rho_+-\rho_-)\,,\nn
\mathbf{j}_s&=\sum_p q_\s n_p \mbf{v}_p\equiv\rho_s \mbf{v}_s\,,
\end{align}
where $n_p=\delta(\mbf{r}-\mbf{r}_p)$ and $\sigma_p=\pm$ for spins up and down. In the hydrodynamic limit, the above equations determine the average particle velocity $\mbf{v}$ and spin velocity $\mbf{v}_s$, which allows us to define four-vectors $j^\mu=(\rho,\rho\mbf{v})$ and $j_s^\mu=(\rho_s,\rho_s\mbf{v}_s)$. Microscopically, the local spin-dependent currents are defined, in the presence of electromagnetic vector potential $\mbf{a}$ and fictitious vector potential $\mathbf{a}_\s$, by 
\ben
m_e\rho_\s\mathbf{v}_\s={\rm Re}\langle\psi^\dag_\s(-i\hbar\bs{\nabla}-\mathbf{a}_\s-e\mbf{a})\psi_\s\rangle\,,
\een
where $e<0$ is the electron charge.

\subsection{Angular and linear momenta}
\label{CEM}

Our Lagrangian \eqref{lagrangian} contains the dynamics of $\mbf{m}(\mbf{r})$ that is coupled to the current. In this regard, we note that the time component of the fictitious gauge potential (\ref{AA}), $a_\s=-\hbar\p_t\varphi(1-\s\cos\theta)/2$, is a Wess-Zumino action that governs the spin-texture dynamics.\cite{volovikJPC87,braunPRB96,tataraPRP08} The variational equation $\mbf{m}\times\de_\mbf{m}L=0$ gives:
\ben
\rho_s( \partial_t  + \mbf{v}_s\cdot\bs{\del} )\mbf{m} + \mbf{m} \times \delta_\mbf{m}F=0\,.
\label{LL}
\een
To derive this equation, we used the spin-density continuity equation (\ref{CE}) and a gauge-independent identity satisfied by the fictitious potentials: their variations with respect to $\mbf{m}$ are given by
\ben
\delta_\mbf{m}a_{\s\mu}(\mbf{m},\p_\mu\mbf{m})=q_\s\mbf{m}\times\partial_\mu\mbf{m}\,,
\label{deltaA}
\een
where
\ben
\delta_\mbf{m}\equiv \frac{\partial} {\partial\mbf{m} } - \sum_\mu \partial_\mu \frac{\p}{\partial (\p_\mu\mbf{m})}\,.
\een
One recognizes that Eq.~\eqref{LL} is the Landau-Lifshitz (LL) equation, in which the spin density precesses about the effective field given explicitly by
\ben
\mathbf{h}\equiv\delta_\mbf{m}F=-A\p_i^2\mbf{m}\,.
\een
Equation~(\ref{LL}) also includes the well-known \emph{reactive} spin torque: $\bs{\tau}=(\mbf{j}_s\cdot\del)\mbf{m}$,\cite{tserkovPRB06md} which is evidently the change in the local spin-density vector due to the spin angular momentum carried by the itinerant electrons. One can formally absorb this spin torque by defining an advective time derivative $D_t\equiv\p_t+\mbf{v}_s\cdot\bs{\nabla}$, with respect to the average spin drift velocity $\mbf{v}_s$.

Equation (\ref{LL}) may be written in a form that explicitly expresses the conservation of angular momentum:\cite{bazaliyPRB98,landauBOOKv9}
\ben
\delt(\rho_s m_i)  +\partial_j \Pi_{ij}=0\,,
\label{angular momentum}
\een
where the angular-momentum stress tensor is defined by
\ben
\Pi_{ij} = \rho_s v_{sj}m_i-A(\mbf{m}\times\p_j\mbf{m})_i\,.
\label{Ptensor}
\een
Notice that this includes both quasiparticle and collective contributions, which stem respectively from the transport and equilibrium spin currents.

The Lorentz force equation for the electrons, Eq.~(\ref{Lorentz Force}), in turn, leads to a continuity equation for the kinetic momentum density.\cite{volovikJPC87} To see this, let us start with the microscopic perspective:
\ben
\p_t(\rho v_i)=\p_t\sum_pn_p\mbf{v}_p=\sum_p\left(\dot{n}_p\mbf{v}_p+n_p\dot{\mbf{v}}_p\right)\,.
\label{mdv}
\een
Using the Lorentz force equation for the second term, we have:
\begin{align}
m_e\sum_pn_p\dot{\mbf{v}}_p&=\sum_p q_\s n_p(e_i+\epsilon^{ijk}b_kv_{p j})=\sum_p q_\s n_pf_{i\mu}v_p^\mu\nn
&\hspace{-1cm}=\rho_s\mbf{m}\cdot(\p_t\mbf{m}\times\p_i\mbf{m})+\rho_sv_{sj}\mbf{m}\cdot(\p_j\mbf{m}\times\p_i\mbf{m})\nn
&\hspace{-1cm}=(\p_i\mbf{m})\cdot(\delta_\mbf{m}F)=-A(\p_i\mbf{m})\cdot(\p_j^2\mbf{m})\,,
\label{pc1}
\end{align}
utilizing Eq.~(\ref{LL}) to obtain the last line. Coarse-graining the first term of Eq.~(\ref{mdv}), in turn, we find:
\ben
\sum_p\dot{n}_p\mbf{v}_p=-\p_j\sum_p\delta(\mbf{r}-\mbf{r}_p)v_{p i}v_{p j}\to-\p_j\sum_\s\rho_\s v_{\s i}v_{\s j}\,.
\label{pc2}
\een
Putting Eqs.~(\ref{pc1}) and (\ref{pc2}) together, we can finally write Eq.~(\ref{mdv}) in the form:
\ben
m_e\p_t(\rho v_i)+\p_j\left(T_{ij}+m_e\sum_\s\rho_\s v_{\s i}v_{\s j}\right)=0\,,
\label{TT}
\een
where
\ben
T_{ij}=A\left[(\p_i\mbf{m})\cdot(\p_j\mbf{m})-\frac{\delta_{ij}}{2}(\p_k\mbf{m})^2\right]
\label{EMtensor}
\een
is the magnetization stress tensor.\cite{volovikJPC87}

A spin-dependent chemical potential $\hat{\mu}=\hat{K}^{-1}\hat{\rho}$ governed by local density and short-ranged interactions can be trivially incorporated by redefining the stress tensor as
\ben
T_{ij}\to T_{ij}+\frac{\delta_{ij}}{2}\hat{\rho}^T\hat{K}^{-1}\hat{\rho}\,.
\label{TK}
\een
In our notation, $\hat{\mu}=(\mu_+,\mu_-)^T$, $\hat{\rho}=(\rho_+,\rho_-)^T$ and $\hat{K}$ is a symmetric $2\times2$ compressibility matrix in spin space, which includes the degeneracy pressure as well as self-consistent exchange and Hartree interactions. In general, Eq.~(\ref{TK}) is valid only for sufficiently small deviations from the equilibrium density.

Using the continuity equations (\ref{CE}), we can combine the last term of Eq.~(\ref{TT}) with the momentum density rate of change:
\ben
\p_t(\rho_\s v_{\s i})+\p_j(\rho_\s v_{\s i}v_{\s j})=\rho_\s(\p_t+\mbf{v}_\s\cdot\bs{\del})v_{\s i}\,,
\label{TTc}
\een
which casts the momentum density continuity equation in the Euler equation form:
\ben
m_e\sum_\s\rho_\s(\p_t+\mbf{v}_\s\cdot\bs{\del})v_{\s i}+\p_jT_{ij}=0\,.
\een
We do not expect such advective corrections to $\p_t$ to play an important role in electronic systems, however. This is in contrast to the advective-like time derivative in Eq.~(\ref{LL}), which is first order in velocity field and is crucial for capturing spin-torque physics.

\subsection{Hydrodynamic free energy}

We will now turn to the Hamiltonian formulation and construct the free energy for our magnetohydrodynamic variables. This will subsequently allow us to develop a nonequilibrium thermodynamic description. The canonical momenta following from the Lagrangian (\ref{lagrangian}) are
\begin{align}
\mbf{p}_p&\equiv\frac{\p L}{\partial\mbf{v}_p}=m_e\mbf{v}_p+\mbf{a}_p\,,\nn
\boldsymbol{\pi}&\equiv\frac{\p\mathcal{L}}{\partial\dot{\mbf{m}}}=\sum_p n_p\frac{\p a_\s}{\partial\dot{\mbf{m}}}=\sum_p n_p a^{\rm mon}_\s(\mbf{m})\,.
\label{canonical momenta}
\end{align}
Notice that for our translationally-invariant system, the total linear momentum
\ben
\mbf{P}\equiv\sum_p\mbf{p}_p+\int d^3r(\boldsymbol{\pi}\cdot\boldsymbol{\nabla})\mbf{m}=m_e\sum_p\mbf{v}_p\,,
\een
where we have used Eq.~(\ref{gauge fields}) to obtain the second equality, coincides with the kinetic momentum (mass current) of the electrons. The latter, in turn, is equivalent to the linear momentum of the original problem of interacting nonrelativistic electrons, in the absence of any real or fictitious gauge fields.  See appendix~\ref{mba}. While $\mbf{P}$ is conserved (as discussed in the previous section and also follows now from the general principles), the canonical momenta of the electrons and the spin-texture field, Eqs.~(\ref{canonical momenta}), are not conserved separately. As was pointed out by Volovik in Ref.~\onlinecite{volovikJPC87}, this explains anomalous properties of the linear momentum associated with the Wess-Zumino action of the spin-texture field: This momentum has neither spin-rotational nor gauge invariance. The reason is that the spin-texture dynamics define only one piece of the total momentum, which is associated with the coherent degrees of freedom. Including also the contribution associated with the incoherent (quasiparticle) background restores the proper gauge-invariant momentum, $\mbf{P}$, which corresponds to the generator of the global translation in the microscopic many-body description.

Performing a Legendre transformation to Hamiltonian as a function of momenta, we find
\begin{align}
H[\mbf{r}_p,\mbf{p}_p;\mbf{m},\bs{\pi}]&=\sum_p\mbf{v}_p\cdot\mbf{p}_p+\int d^3r\dot{\mbf{m}}\cdot \boldsymbol{\pi}-L\nn
&=\sum_p \frac{(\mbf{p}_p-\mbf{a}_\s)^2}{2m_e}+\frac{A}{2}\int d^3r(\p_i\mbf{m})^2\nn
&\equiv E+F\,,
\label{hamiltonian}
\end{align}
where $E$ is the kinetic energy of electrons and $F$ is the exchange energy of the magnetic order. As could be expected, $E$ is the familiar single-particle Hamiltonian coupled to an external vector potential. According to a Hamilton's equation, the velocity is conjugate to the canonical momentum: $\mbf{v}_p=\partial H/\p\mbf{p}_p$. We note that explicit dependence on the spin-texture dynamics dropped out because of the special property of the gauge fields: $\dot{\mbf{m}}\cdot\partial_{\dot{\mbf{m}}}a_\s=a_\s$.  Furthermore, according to Eq.~\eqref{deltaA}, we have $\mbf{m}\times\delta_\mbf{m}E=(\mbf{j}_s\cdot\bs{\del})\mbf{m}$, so the LL Eq.~\eqref{LL} can be written in terms of the Hamiltonian \eqref{hamiltonian} as\cite{tserkovPRB09md}
\ben
\rho_s \dot{\mbf{m}}+\mbf{m}\times\delta_\mbf{m}H=0\,.
\een

So far, we have included in the spin-texture equation only the piece coupled to the itinerant electron degrees of freedom. The purely magnetic part is tedious to derive directly and we will include it in the usual LL phenomenology.\cite{landauBOOKv9} To this end, we redefine
\ben
F[\mbf{m}(\mbf{r})]\to F+F^\prime\,,
\een
by adding an additional magnetic free energy $F^\prime[\mbf{m}(\mbf{r})]$, which accounts for magnetostatic interactions, crystalline anisotropies, coupling to external fields, as well as energy associated with localized $d$ or $f$ orbitals.\cite{nEnergy} Then the total free energy (Hamiltonian) is $H=E+F$, and  we in general define the effective magnetic field as the thermodynamic conjugate of $\mbf{m}$: $\mbf{h}\equiv\de_\mbf{m}H$ .  The LL equation then becomes
\ben
\varrho_s\dot{\mbf{m}}+\mbf{m}\times\mbf{h}=0\,,
\label{LL3}
\een
where $\varrho_s$ is the total effective spin density.  To enlarge the scope of our phenomenology, we allow the possibility that $\varrho_s\neq\rho_s$. For example, in the $s-d$ model, an extra spin density comes from the localized $d$-orbital electrons. Microscopically, $\varrho_s\p_t\mbf{m}$ term in the equation of motion stems from the Wess-Zumino action generically associated with the total spin density.

In the following, it may sometimes be useful to separate out the current-dependent part of the effective field, and write the purely magnetic part as $\mbf{h}_m\equiv\de_\mbf{m}F$, so that
\ben
\mbf{h}=\mbf{h}_m-\mbf{m}\times(\mbf{j}_s\cdot\bs{\del})\mbf{m}
\een
and Eq.~(\ref{LL3}) becomes:
\ben
\varrho_s\dot{\mbf{m}} + (\mbf{j}_s\cdot\bs{\del})\mbf{m}+\mbf{m}\times\mbf{h}_m=0\,.
\label{LLL}
\een
For completeness, let is also write the equation of motion for the spin-$\s$ acceleration:
\begin{align}
m_e(\p_t+\mbf{v}_\s\cdot\bs{\del})v_{\s i}&=q_\s[\mbf{m}\cdot(\p_t\mbf{m}\times\p_i\mbf{m})\nn
&\hspace{-0.5cm}+v_{\s j}\mbf{m}\cdot(\p_j\mbf{m}\times\p_i\mbf{m})]-\bs{\del}\mu_\s\,,
\label{EE}
\end{align}
retaining for the moment the advective correction to the time derivative on the left-hand side and reinserting the force due to the spin-dependent chemical potential, $\hat{\mu}=\hat{K}^{-1}\hat{\rho}$. These equations constitute the coupled reactive equations for our magneto-electric system. The Hamiltonian (free energy) in terms of the collective variables is (including the elastic compression piece)
\begin{align}
H[\rho_\s,\mbf{p}_\s;\mbf{m}]=&\sum_\s \int d^3r\rho_\s\frac{(\mbf{p}_\s-\mbf{a}_\s)^2}{2m_e}\nn
&+\frac{1}{2}\int d^3r\hat{\rho}^T\hat{K}^{-1}\hat{\rho}+F[\mbf{m}]\,,
\label{FH}
\end{align}
where $\mbf{p}_\s=m_e\mbf{v}_\s+\mbf{a}_\s$ is the spin-dependent momentum that is locally averaged over individual particles.

\subsection{Conservation of energy}

So far, our hydrodynamic equations are reactive, so that the energy (\ref{FH}) must be conserved: $P\equiv\dot{H}=\dot{E}+\dot{F}=0 $. The time derivative of the electronic energy $E$ is
\begin{align}
\dot{E}&=\int d^3r\sum_\s\left[m_e\rho_\s\mbf{v}_\s\dot{\mbf{v}}_\s+\dot{\rho}_\s\left(\frac{m_ev_\s^2}{2}+\mu_\s\right)\right]\nn
&=\int d^3r\sum_\s\left[m_e\rho_\s v_{\s j}\dot{v}_{\s j}-\p_j(\rho_\s v_{\s j})\left(\frac{m_e v_\s^2}{2}+\mu_\s\right)\right]\nn
&=\int d^3r\sum_\s\rho_\s v_{\s j}\left[m_e\left(\p_t+\mbf{v}_\s\cdot\bs{\del}\right)v_{\s j}+\p_j\mu_\s\right]\nn
&=\int d^3r\sum_\s q_\s\rho_\s\mbf{v}_\s\cdot(\mbf{e}+\mbf{v}_\s\times\mbf{b})\nn
&=\int d^3r\sum_\s q_\s\rho_\s\mbf{v}_\s\cdot\mbf{e}=\int d^3r\mbf{j}_s\cdot\mbf{e}\,.
\label{work}
\end{align}
The change in the spin-texture energy is given, according to Eq.~(\ref{LLL}), by
\begin{align}
\dot{F}&= \int d^3r\dot{\mbf{m}}\cdot\de_\mbf{m}F=\int d^3r\dot{\mbf{m}}\cdot\mbf{h}_m\nn
&=\int d^3r\dot{\mbf{m}}\cdot\left[\varrho_s\mbf{m}\times\dot{\mbf{m}}+ \mbf{m}\times (\mbf{j}_s\cdot\bs{\del})\mbf{m})\right]\nn
&=-\int d^3r\mbf{j}_s\cdot\mbf{e}\,.
\end{align}
The total energy is thus evidently conserved, $P=0$. When we calculate dissipation in the rest of the paper, we will omit these terms which cancel each other. The total energy flux density is evidently given by
\ben
\mbf{Q}=\sum_\s\rho_\s\left(\frac{m_ev^2_\s}{2}+\mu_\s\right)\mbf{v}_\s\,.
\een

\section{Dissipation}
\label{DISS}

Having derived from first principles the reactive couplings in our magneto-electric system, summed up in Eqs.~(\ref{LLL})-(\ref{FH}), we will proceed to include the dissipative effects phenomenologically. Let us focus on the linearized limit of small deviations from equilibrium (which may be spin textured), so that the advective correction to the time derivative in the Euler Eq.~(\ref{EE}), which is quadratic in the velocity field, can be omitted. To eliminate the quasiparticle spin degree of freedom, let us, furthermore, treat halfmetallic ferromagnets, so that $\rho=\rho_+$ and $\rho_s=q\rho$, where $q=\hbar/2$ is the electron's spin.\cite{nTwo} From Eq.~(\ref{EE}), the equation of motion for the local (averaged) canonical momentum is:\cite{nCC}
\ben
\dot{\mbf{p}}=\frac{q}{\rho}\mbf{j}\times \mathbf{b}-\bs{\del}\mu\,,
\label{canonical}
\een
in a gauge where $a_\s=0$, so that $\dot{\mbf{p}}=m_e\dot{\mbf{v}}-q\mbf{e}$.\cite{nEM} $\mu=\rho/K$. The Lorentz force due to the applied (real) electromagnetic fields  can be added in the obvious way to the right-hand side of Eq.~(\ref{canonical}). Note that since we are now interested in linearized equations close to equilibrium, $\rho$ in Eq.~(\ref{canonical}) can be approximated by its (homogeneous) equilibrium value.

Introducing relaxation through a phenomenological damping constant (Drude resistivity)
\ben
\gamma={m_e\over \rho\tau}\,,
\een
where $\tau$ is the collision time, expressing the fictitious magnetic field in terms of the spin texture, Eq.~\eqref{canonical} becomes:
\ben
\dot{p}_i=-\frac{q}{\rho}(\mbf{m}\times\p_i\mbf{m})\cdot(\mbf{j}\cdot\bs{\del})\mbf{m}-\p_i\mu-\gamma j_i\,.
\label{canonical2}
\een

Adding the phenomenological Gilbert damping\cite{gilbertIEEEM04} to the magnetic Eq.~(\ref{LL3}) gives the Landau-Lifshitz-Gilbert equation:
\ben
\varrho_s(\dot{\mbf{m}}+\alpha\mbf{m}\times\dot{\mbf{m}})= \mbf{h}\times\mbf{m}\,,
\label{LLG}
\een
where $\al$ is the damping constant. Eqs.~\eqref{canonical2} and \eqref{LLG}, along with the continuity equation, $\dot{\rho}=-\bs{\del}\cdot\mbf{j}$, are the near-equilibrium thermodynamic equations for $(\rho,\mbf{p},\mbf{m})$ and their respective thermodynamic conjugates $(\mu,\mbf{j},\mbf{h})=(\delta_\rho H,\delta_\mbf{p}H,\delta_\mbf{m}H)$. This system of equations of motion may be written formally as 
\ben
\delt \left( \begin{array}{c}\rho\\\mbf{p}\\ \mbf{m}\end{array}\right)= \widehat{\Gamma}[\mbf{m}(\mbf{r})]\left( \begin{array}{c}\mu\\\mbf{j}\\ \mbf{h}\end{array}\right)\,.
\label{rpm}
\een
The matrix $\hat{\Gamma}$ depends on the equilibrium spin texture $\mbf{m}(\mbf{r})$. By the Onsager reciprocity principle, $\Gamma_{ij}[\mbf{m}]=s_is_j\Gamma_{ji}[-\mbf{m}]$, where $s_i=\pm$ if the $i$th variable is even (odd) under time reversal.

In the quasistationary description of a nonequilibrium thermodynamic system, the entropy $S[\rho,\mbf{p},\mbf{m}]$ is formally regarded as a functional of the instantaneous thermodynamic variables, and the probability of a given configuration is proportional to $e^{S/k_B}$. If the heat conductance is high and the temperature $T$ is uniform and constant, the instantaneous rate of dissipation $P=T\dot{S}$ is given by the rate of change in the free energy, $P=\dot{H}=\int d^3r\mcal{P}$:
\ben
\mathcal{P}=-\mu\dot{\rho}-\mbf{h}\cdot\dot{\mbf{m}}-\mbf{j}\cdot\dot{\mbf{p}}=\al\varrho_s\dot{\mbf{m}}^2 +\gamma\mbf{j}^2\,,
\een
where we used Eq.~\eqref{canonical2} and expressed the effective field $\mbf{h}$ as a function of $\dot{\mbf{m}}$ by taking $\mbf{m}\times$ of Eq.~\eqref{LLG}:
\ben
\mbf{h}=\varrho_s\mbf{m} \times \dot{\mbf {m}} -\al\varrho_s\dot{\mbf{m}}\,.
\label{inverseLLG}
\een
Notice that the fictitious magnetic field $\mbf{b}$ does not contribute to dissipation because it does not do work.

So far, there is no dissipative coupling between the current and the spin-texture dynamics, and the macroscopic equations obey the global time-reversal symmetry. However, we know that dissipative couplings exists due to the misalignment of the electron's spin with the collective spin texture and spin-texture resistivity.\cite{marrowsAP05,tserkovPRB06md} Following Ref.~\onlinecite{tserkovPRB09md}, we add these well-known effects phenomenologically by making an expansion in the equations of motion to linear order in the nonequilibrium quantities $\dot{\mbf{m}}$ and $\mbf{j}$. To limit the number of terms one can write down, we will only add terms that are spin-rotationally invariant and isotropic in real space (which disregards, in particular, such effects as the angular magnetoresistance and the anomalous Hall effect). To second order in the spatial gradients of $\mbf{m}$, there are only three dissipative phenomenological terms with couplings $\eta$, $\eta^\prime$, and $\beta$ consistent with the above requirements, which could be added to the right-hand side of Eq.~(\ref{canonical2}).\cite{nViscosity} The momentum equation becomes:
\begin{align}
\dot{p}_i=&-\frac{q}{\rho}(\mbf{m}\times\p_i\mbf{m})\cdot(\mbf{j}\cdot\bs{\del})\mbf{m}-\p_i\mu-\gamma j_i\nn
-&\eta(\p_k\mbf{m})^2j_i-\eta^\prime\partial_i\mbf{m}\cdot(\mbf{j}\cdot\bs{\del})\mbf{m}-q\beta\dot{\mbf{m}}\cdot\partial_i\mbf{m}\,.
\label{canonical3}
\end{align}
It is known that the ``$\beta$ term" comes from a misalignment of the electron spin with the collective spin texture, and the associated dephasing.  It is natural to expect thus that the dimensionless parameter $\beta\sim\hbar/\tau_s\D$, where $\tau_s$ is a characteristic spin-dephasing time.\cite{tserkovPRB06md} The ``$\eta$ terms" evidently describe texture-dependent resistivity, which is anisotropic with respect to the gradients in the spin texture along the local current density. Such term are also naturally expected, in view of the well-known giant-magnetoresistance effect,\cite{baibichPRL88} in which noncollinear magnetization results in electrical resistance. The microscopic origin of this term is due to spin-texture misalignment, which modifies electron scattering.

The total spin-texture-dependent resistivity can be put into a tensor form:
\begin{align}
\gamma_{ij}[\mbf{m}]=&\delta_{ij}\left[\gamma+\eta(\p_k\mbf{m})^2\right]+\eta^\prime\p_i\mbf{m}\cdot\p_j\mbf{m}\nn
&+\frac{q}{\rho}\mbf{m}\cdot(\partial_i\mbf{m}\times\p_j\mbf{m})\,.
\label{damping}
\end{align}
The last term due to fictitious magnetic field gives a Hall resistivity. Note that $\hat{\gamma}[\mbf{m}]=\hat{\gamma}^T[-\mbf{m}]$, consistent with the Onsager theorem.  We can finally write Eq.~\eqref{canonical2} as:
\ben
\dot{p}_i=-\gamma_{ij}[\mbf{m}]j_j -\p_i\mu-q\beta\dot{\mbf{m}}\cdot \partial_i\mbf{m}\,.
\label{canonical4}
\een
As was shown in Ref.~\onlinecite{tserkovPRB09md}, since the Onsager relations require that $\widehat\Gamma[\mbf{m}]= \widehat\Gamma[-\mbf{m}]^T$ within the current/spin-texture fields sector, there must be a counterpart to the $\beta$ term above in the magnetic equation, which is the well-known dissipative ``$\beta$ spin torque:"
\ben
\varrho_s(\dot{\mbf{m}}+\alpha\mbf{m}\times\dot{\mbf{m}})=\mbf{h}\times\mbf{m}-q\beta \mbf{m}\times(\mbf{j}\cdot\bs{\del})\mbf{m}\,.
\label{LLGbeta}
\een

The total dissipation $\mcal{P}$ is now given by
\begin{align}
\mcal{P}=&\al\varrho_s\dot{\mbf{m}}^2+2q\beta\dot{\mbf{m}}\cdot(\mbf{j}\cdot\bs{\del})\mbf{m}+\left[\gamma+\eta(\p_k\mbf{m})^2\right]j^2\nn
&+\eta^\prime[(\mbf{j}\cdot\bs{\del})\mbf{m}]^2\nn
=&\al\varrho_s\left[\dot{\mbf{m}}+\frac{q\beta}{\al\varrho_s}(\mbf{j}\cdot\bs{\del})\mbf{m}\right]^2+\left[\gamma+\eta(\p_k\mbf{m})^2\right]j^2\nn
&+\left(\eta^\prime-\frac{(q\beta)^2}{\alpha\varrho_s}\right)[(\mbf{j}\cdot\bs{\del})\mbf{m}]^2\,.
\label{diss}
\end{align}
The second law of thermodynamics requires the total dissipation to be positive, which puts some constraints on the allowed values of the phenomenological parameters. We can easily notice, however, that the dissipation (\ref{diss}) is guaranteed to be positive-definite if 
\ben
\eta+\eta^\prime\geq\frac{(q\beta)^2}{\al\varrho_s}\,,
\een
which may serve as an estimate for the spin-texture resistivity due to spin dephasing. This is consistent with the microscopic findings of Ref.~\onlinecite{WicklesPRB09}.

\section{Thermal Noise}
\label{TN}

At finite temperature, thermal agitation causes fluctuations of the current and spin texture, which are correlated due to their coupling.  A complete description requires that we supplement the stochastic equations of motion with the correlators for these fluctuations. It is convenient to regard these fluctuations as being due to the stochastic Langevin ``forces" $(\delta\mu,\delta\mbf{j},\delta\mbf{h})$ on the right-hand side of Eq.~(\ref{rpm}).  The complete set of finite-temperature hydrodynamic equations thus becomes:
\begin{align}
\dot{\rho}&=-\bs{\del}\cdot\tilde{\mbf{j}}\,,\nn
\dot{\mathbf{p}}+q\beta\dot{m}_i\bs{\nabla}m_i&=-\hat{\gamma}[\mbf{m}]\tilde{\mbf{j}}-\bs{\nabla}\tilde{\mu}\,,\nn
\varrho_s(1+\alpha\mbf{m}\times)\dot{\mbf{m}}&=\tilde{\mbf{h}}\times\mbf{m}-q\beta\mbf{m}\times(\tilde{\mbf{j}}\cdot\bs{\del})\mbf{m}\,.
\label{stochastic}
\end{align}
where $(\tilde{\mu},\tilde{\mbf{j}},\tilde{\mbf{h}})=(\mu+\delta\mu,\mbf{j}+\delta\mbf{j},\mbf{h}+\delta\mbf{h})$. The simplest (while possibly not most realistic) case corresponds to a highly compressible fluid, such that $K\to\infty$. In this limit, $\mu=\rho/K\to0$ and the last two equations completely decouple from the first, continuity equation. In the remainder of this section, we will focus on this special case. The correlations of the stochastic fields are given by the symmetric part of the inverse matrix $\widehat{\Upsilon}=-\widehat{\Gamma}^{-1}$,\cite{landauBOOKv5} which is found by inverting Eq.~(\ref{stochastic}) (reduced now to a system of two equations):
\begin{align}
\tilde{\mbf{j}}&=-\hat{\gamma}^{-1}\left(\dot{\mbf{p}}+q\beta\dot{m}_i\bs{\del} m_i\right)\,,\nn
\tilde{\mbf{h}}&=\varrho_s\mbf{m}\times\dot{\mbf{m}}-\al\varrho_s\dot{\mbf{m}}-q\beta(\tilde{\mbf{j}}\cdot\bs{\del})\mbf{m}\,.
\label{inverseEOM}
\end{align}
Writing formally these equations as (after substituting $\tilde{\mbf{j}}$ from the first into the second equation)
\ben
\left(\begin{array}{c}\tilde{\mbf{j}}\\ \tilde{\mbf{h}}\end{array}\right)=-\widehat{\Upsilon}[\mbf{m}(\mbf{r})]\left( \begin{array}{c}\dot{\mbf{p}}\\ \dot{\mbf{m}}\end{array}\right)\,,
\een
we immediately read out for the matrix elements $\widehat{\Upsilon}(\mbf{r},\mbf{r}^\prime)=\widehat{\Upsilon}(\mbf{r})\delta(\mbf{r}-\mbf{r}^\prime)$:
\begin{align}
\Upsilon_{j_i,j_{i^\prime}}(\mbf{r})=&(\hat{\gamma}^{-1})_{ii^\prime}\,,\nn
\Upsilon_{j_i,h_{i^\prime}}(\mbf{r})=&q\beta(\hat{\gamma}^{-1})_{ik}\p_k m_{i'}\,,\nn
\Upsilon_{h_{i^\prime},j_i}(\mbf{r})=&-q\beta(\hat{\gamma}^{-1})_{ki}\p_k m_{i'}\,,\nn
\Upsilon_{h_i,h_{i^\prime}}(\mbf{r})=&\al\varrho_s\delta_{ii^\prime}+\varrho_s\e^{ii'k}m_k\nn
&-(q\beta)^2(\p_k m_i)(\hat{\gamma}^{-1})_{kk^\prime}(\p_{k^\prime}m_{i^\prime})\,.
\end{align}
According to the fluctuation-dissipation theorem, we symmetrize $\widehat{\Upsilon}$ to obtain the classical Langevin correlators:\cite{landauBOOKv5}
\begin{align}
\langle\delta j_i(\mbf{r},t)\delta j_{i^\prime}(\mbf{r}^\prime,t^\prime)\rangle/\mcal{T}&=g_{ii^\prime}\,,\nn
\langle\delta j_i(\mbf{r},t)\delta h_{i^\prime}(\mbf{r}^\prime,t^\prime)\rangle/\mcal{T}&=q\beta g^\prime_{ik}\p_km_{i^\prime}\,,\nn
\langle\delta h_i(\mbf{r},t)\delta h_{i^\prime}(\mbf{r}^\prime,t^\prime)\rangle/\mcal{T}&=\alpha\varrho_s\delta_{ii'}\nn
&\hspace{-1cm}-(q\beta)^2g_{kk^\prime}(\p_km_i)(\p_{k^\prime}m_{i^\prime})\,,
\label{langevin}
\end{align}
where $\mcal{T}=2k_BT\delta(\mbf{r}-\mbf{r}')\delta(t-t^\prime)$ and
\begin{equation}
\hat{g}=[\hat{\gamma}^{-1}+(\hat{\gamma}^{-1})^T]/2\,,\,\,\,\hat{g}^\prime=[\hat{\gamma}^{-1}-(\hat{\gamma}^{-1})^T]/2
\end{equation}
are, respectively, the symmetric and antisymmetric parts of the conductivity matrix $\hat{\gamma}^{-1}$. The short-ranged, $\delta$-function character of the noise correlations in space stems from the assumption of high electronic compressibility. Contrast this to the results of Ref.~\onlinecite{tserkovPRB09md} for incompressible hydrodynamics. A presence of long-ranged Coulombic interactions and plasma modes would also give rise to nonlocal correlations. These are absent in our treatment, which disregards ordinary electromagnetic phenomena.

Focusing on the microwave frequencies $\omega$ characteristic of ferromagnetic dynamics, it is most interesting to consider the regime where $\omega\ll\tau^{-1}$. This means that we can employ the drift approximation for the first of Eqs.~(\ref{inverseEOM}):
\ben
\dot{p}_i=m_e\dot{v}_i-q e_i\approx-qe_i=q\dot{\mbf{m}}\cdot(\mbf{m}\times\p_i\mbf{m})\,.
\label{da}
\een
Substituting this $\dot{\mbf{p}}$ in Eq.~(\ref{inverseEOM}), we can easily find a closed stochastic equation for the spin-texture field:
\begin{align}
&\varrho_s(1+\al \mbf{m}\times)\dot{\mbf{m}}+\mbf{m}\times\tensor{\tau}\dot{\mbf{m}}=(\mbf{h}_m+\de\underline{\mbf{h}})\times\mbf{m}\,,
\label{LLGbeta2}
\end{align}
where we have defined the ``spin-torque tensor"
\begin{align}
\tensor{\tau}=q^2(\hat{\gamma}^{-1})_{kk^\prime}&\left( \mbf{m}\times\p_k\mbf{m}-\beta\p_k\mbf{m}\right)\nn
&\otimes\left(\mbf{m}\times\p_{k^\prime}\mbf{m}+\beta\p_{k^\prime}\mbf{m}\right)\,.
\end {align}
The antisymmetric piece of this tensor modifies the effective gyromagnetic ratio, while the more interesting symmetric piece determines the additional nonlocal Gilbert damping:
\ben
\tensor{\alpha}=\frac{\tensor{\tau}+\tensor{\tau}^T}{2\varrho_s}=\frac{q^2}{\varrho_s}\tensor{G}\,,
\een
where
\begin{align}
\tensor{G}=&g_{kk'}\left[(\mbf{m}\times\p_k\mbf{m})\otimes(\mbf{m}\times\p_{k^\prime}\mbf{m})-{\beta^2}\partial_k\mbf{m}\otimes\partial_{k^\prime}\mbf{m}\right]\nn
&+\beta g^{\prime}_{kk^\prime}\left[(\mbf{m}\times\p_k\mbf{m})\otimes\p_{k^\prime}\mbf{m}-\p_k\mbf{m}\otimes(\mbf{m}\times\p_{k^\prime}\mbf{m})\right]\,.
\label{Gtensor}
\end{align}
In obtaining Eq.~(\ref{LLGbeta2}) from Eqs.~(\ref{inverseEOM}), we have separated the reactive spin torque out of the effective field: $\mbf{h}=\mbf{h}_m-q\mbf{m}\times(\mbf{j}\cdot\bs{\del})\mbf{m}$. (The remaining piece $\mbf{h}_m$ thus reflects the purely magnetic contribution to the effective field.) The total stochastic magnetic field entering Eq.~(\ref{LLGbeta2}),
\begin{equation}
\delta\underline{\mbf{h}}=\delta\mbf{h}+q\mbf{m}\times(\delta\mbf{j}\cdot\bs{\del})\mbf{m}\,,
\end{equation}
captures both the usual magnetic Brown noise\cite{brownPR63} $\delta\mbf{h}$ and the Johnson noise spin-torque contribution\cite{forosPRB08} $\delta\mbf{h}_J=q\mbf{m}\times(\delta\mbf{j}\cdot\bs{\del})\mbf{m}$ that arises due to the substitution $\mbf{j}=\tilde{\mbf{j}}-\delta\mbf{j}$ in the reactive spin torque $q(\mbf{j}\cdot\bs{\del})\mbf{m}$. Using correlators (\ref{langevin}), it is easy to show that the total effective field fluctuations $\delta\underline{\mbf{h}}$ are consistent with the nonlocal effective Gilbert damping tensor (\ref{Gtensor}), in accordance with the fluctuation-dissipation theorem applied directly to the purely magnetic Eq.~(\ref{LLGbeta2}).

To the leading, quadratic order in spin texture, we can replace $g_{kk^\prime}\to\delta_{kk^\prime}/\gamma$ and $g_{kk^\prime}^\prime\to0$ in Eq.~(\ref{Gtensor}). This additional texture-dependent nonlocal damping (along with the associated magnetic noise) is a second-order effect, physically corresponding to the backaction of the magnetization dynamics-driven current on the spin texture.\cite{tserkovPRB09md} It should be noted that in writing the modified LLG equation (\ref{LLGbeta}), we did not systematically expand it to include the most general phenomenological terms up to the second order in spin texture. We have only included extra spin-torque terms, which are required by the Onsager symmetry with Eq.~(\ref{canonical3}). The second-order Gilbert damping (\ref{Gtensor}) was then obtained by solving Eqs.~(\ref{canonical3}) and (\ref{LLGbeta}) simultaneously. (Cf. Refs.~\onlinecite{tserkovPRB09md,zhangPRL09}.) This means in particular, that this procedure does not capture second-order Gilbert damping effects whose physical origin is unrelated to the longitudinal spin-transfer torque physics studied here. One example of that is the transverse spin-pumping induced damping discussed in Refs.~\onlinecite{hankiewiczPRB08}.

\section{Examples}
\label{EX}

\subsection{Rigidly spinning texture}
 
\begin{figure}[htbp]
\centerline{\includegraphics[width=\linewidth] {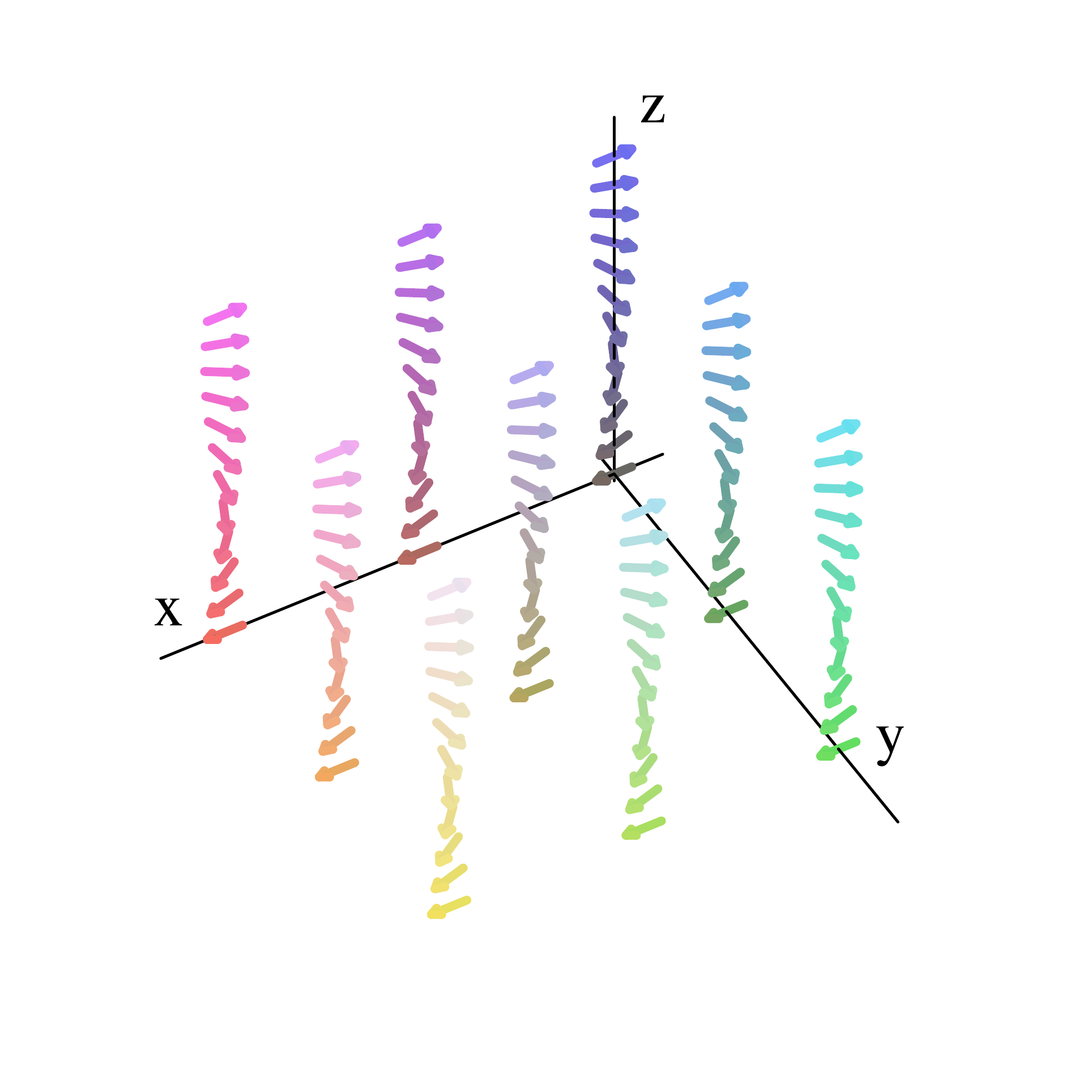}}
\caption{(Color online) The transverse magnetic helix, $\p_z\mbf{m}=\kappa\mbf{\hat{z}}\times\mbf{m}$, with texture-dependent anisotropic resistivity (\ref{ran}). We assume here translational invariance in the transverse ($xy$) directions. Spinning this helix about the vertical $z$ axis generates the dissipative electromotive forces $f_z^{(\beta)}$, which is spatially uniform and points everywhere along the $z$ axis. A magnetic spiral, $\p_z\mbf{m}=\kappa\bs{\hat{\varphi}}\times\mbf{m}=\kappa\bs{\hat{\theta}}$, spinning around the $z$ axis, on the other hand, produces a purely reactive electromotive force $e_z$, as discussed in the text, which is oscillatatory in space along the $z$ axis.}
\label{helix}
\end{figure}

To illustrate the $\eta$ resistivity terms in the electron's equation of motion \eqref{canonical3}, we first consider 1D textures. Take, for example, the case of a 1D spin helix $\mbf{m}(z)$ along the $z$ axis, whose spatial gradient profile is given by $\p_z\mbf{m}=\kappa\mbf{\hat{z}}\times\mbf{m}$, where $\kappa$ is the wave vector of the spatial rotation and $\mbf{m}\perp\mbf{\hat{z}}$. See Fig.~\ref{helix}. It gives anisotropic resistivity in the $xy$ plane, $r^{(\eta)}_\perp$, and along the $z$ direction, $r^{(\eta)}_\parallel$:
\ben
r^{(\eta)}_\perp=\eta(\p_z\mbf{m})^2=\eta\kappa^2\,,\,\,\,r^{(\eta)}_\parallel=(\eta+\eta^\prime)\kappa^2\,.
\label{ran}
\een

The fictitious electric field and dissipative $\beta$ force require magnetic dynamics. A general texture globally rotating clockwise in spin space in the $xy$ plane according to $\dot{\mbf{m}}=-\omega \mbf{\hat{z}}\times\mbf{m}$ (which may be induced by applying a magnetic field along the $z$ direction) generates an electric field
\begin{align}
e_i&=(\mbf{m}\times\dot{\mbf{m}})\cdot\p_i\mbf{m}=-\omega(\mbf{m}\times\mbf{\hat{z}}\times\mbf{m})\cdot\p_i\mbf{m}\nn
&=-\omega\p_im_z=-\omega\p_i\cos\theta
\label{ei}
\end{align}
and a $\beta$ force
\begin{align}
f^{(\beta)}_i&=-\beta\dot{\mbf{m}}\cdot\p_i\mbf{m}=\beta\omega\mbf{\hat{z}}\cdot(\mbf{m}\times\p_i\mbf{m})\nn
&=\beta\omega\sin^2{\theta}\p_i\varphi\,,
\label{fi}
 \end{align}
 where  $(\theta,\varphi)$  denote the position-dependent spherical angles parametrizing the spin texture. The reactive force (\ref{ei}) has a simple interpretation of the gradient of the Berry-phase\cite{berryPRSLA84} accumulation rate [which is locally determined by the solid angle subtended by $\mbf{m}(t)$]. In the case of the transverse helix discussed above, $\theta=\pi/2$, $\varphi=\kappa z-\omega t$, so that $e_z=0$  while $f_z^{(\beta)}=-\beta\omega\kappa$ is finite.

As an example of a dynamical texture that does not generate $\mbf{f}^{(\beta)}$ while producing a finite $\mbf{e}$, consider a spin spiral along the $z$ axis, described by $\p_z\mbf{m}=\kappa\bs{\hat{\varphi}}\times\mbf{m}=\kappa\bs{\hat{\theta}}$, and rotating in time in the manner described above.  It is clear geometrically that the change in the spin texture in time is in a direction orthogonal to its gradients in space.  Specifically, $\theta=\kappa z$, $\varphi=-\omega t$, so that $f^{(\beta)}_z=0$ while the electric field is oscillatory, $e_z=\omega\kappa\sin\theta$.

\subsection{Rotating spin textures}

We show here that a vortex rotating about its core in orbital space generates a current circulating around its core, as well as a current going radially with respect to the core.   Consider a spin texture with a time dependence corresponding to the real-space rotation  clockwise in the $xy$ plane around the origin, such that $\mbf{m}(\mbf{r},t)=\mbf{m}(\mbf{r}(t),0)$ with $\dot{\mbf{r}}= \omega\mbf{\hat{z}}\times\mbf{r}=\omega r \bs{\hat{\phi}}$, where we use  polar coordinates $(r,\phi)$ on the plane normal to the $z$ axis in real space [to be distinguished from the spherical coordinates $(\theta,\varphi)$ that parametrize $\mbf{m}$ in spin space], we have  
 \ben
 \dot{\mbf{m}}=(\dot{\mbf{r}}\cdot\bs{\del})\mbf{m}=\omega \p_\phi\mbf{m}\,.
 \een
For $\mbf{m}(r,\phi)$ in polar coordinates, the components of the electric field are,
 \ben
e_r= \omega \mbf{m}\cdot (\p_\phi\mbf{m} \times\p_r\mbf{m})\,,\,\,\,e_\phi=0\,,
\een
while the components of the $\beta$ force are
\ben
f^{(\beta)}_r=-\beta \omega (\p_r\mbf{m})\cdot(\p_\phi\mbf{m})\,,\,\,\,f^{(\beta)}_\phi=-\beta \omega\frac{(\p_\phi\mbf{m})^2}{r}\,.
\een
In order to find the fictitious electromagnetic fields, we need to calculate the following tensors (which depend on the instantaneous spin texture):
\begin{align}
b_{ij}&\equiv \mbf{m}\cdot(\p_i\mbf{m}\times\p_j\mbf{m})=\sin\theta (\p_i\theta \p_j\varphi-\p_j\theta \p_i\varphi)\,, \nn
d_{ij}&\equiv\p_i\mbf{m}\cdot\p_j\mbf{m}=\p_{i}\theta\p_j\theta+\sin^2\theta \p_{i}\varphi\p_j\varphi\,.
\label{tensors}
 \end {align}

\begin{figure}[htbp]
\centerline{\includegraphics[width=0.7\linewidth]{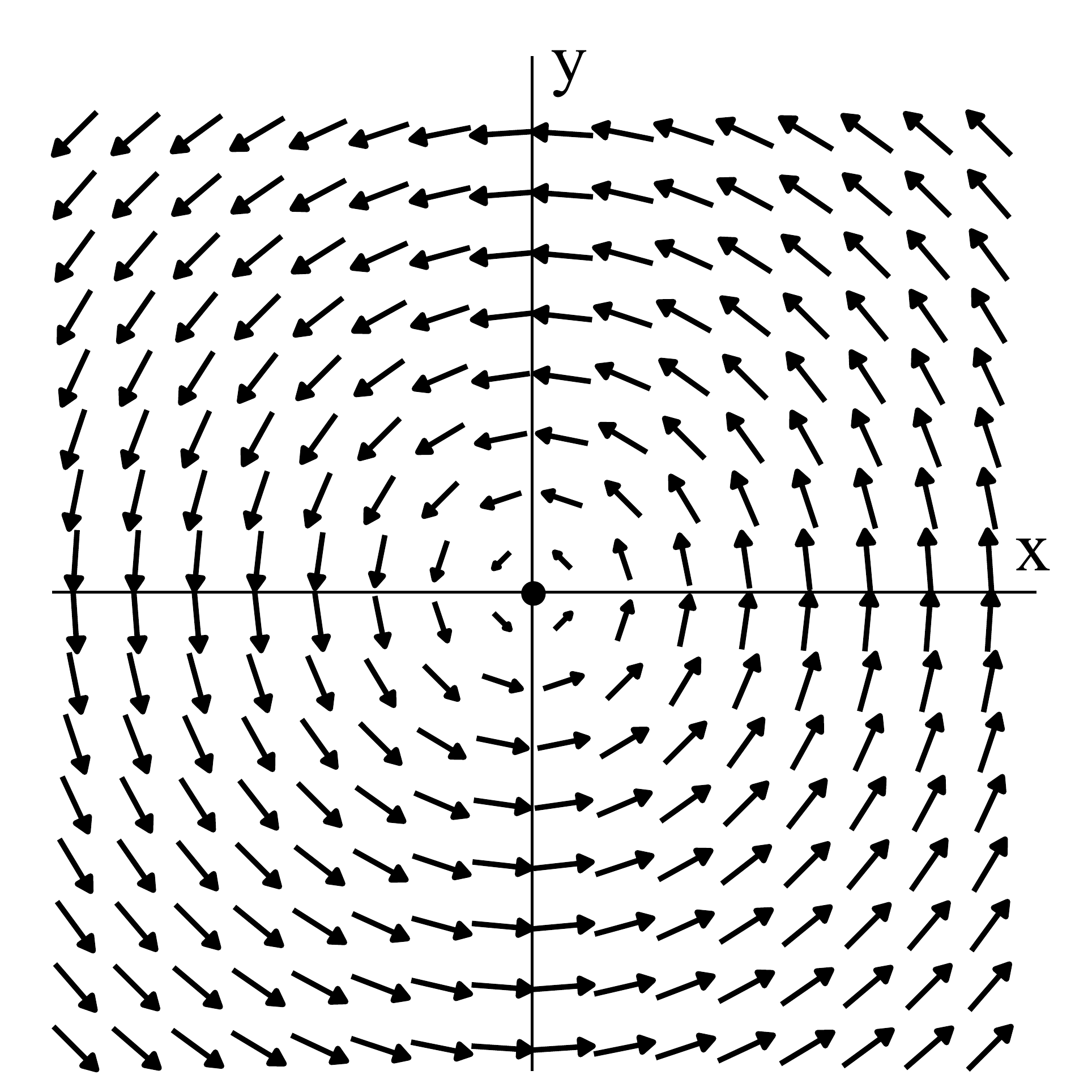}}
\caption{Positive-polarity magnetic vortex configuration projected on the $xy$ plane. $\mbf{m}$ has a positive (out-of-plane) $z$ component near the vortex core. Rotating this vortex about the origin in real space generates the current in the $xy$ plane shown in Fig.~\ref{current}.}
\label{vortex}
\end{figure}

As an example, consider a vortex centered at the origin in the $xy$ plane with winding number 1 and positive polarity, as shown in Fig.~\ref{vortex}.  Its angular coordinates are given by
\ben
\varphi=(\phi+\omega t)+\frac{\pi}{2}\,,\,\,\, \theta=\theta (r)\,,
\een 
where $\phi=\arg(\mbf{r})$ and  $\theta$ is a rotationally invariant function such that $\theta\to0$ as $r\to0$ and $\theta\to\pi/2$ as $r\to\infty$.  Evaluating the tensors in equation \eqref{tensors} for this vortex in polar coordinates gives $d_{rr}= (\p_r\theta)^2$, $d_{\phi\phi}=(\sin\theta/r)^2$, $d_{r\phi}=0$, and $ b_{r\phi}=-(\p_r\cos\theta)/r$.
 The radial electric field is then given by
  \ben
 e_r=-\omega rb_{r\phi}=\omega \p_r\cos\theta\,.
\label{er}
\een
The $\beta$ force is in the azimuthal direction:
 \ben
  f^{(\beta)}_r=0\,,\,\,\, f^{(\beta)}_\phi=-\beta\omega rd_ {\phi\phi}=-\beta\omega\frac{\sin^2\theta}{r}\,.
 \label{fr}
 \een
We can interpret this force as the spin texture ``dragging"  the current along its direction of motion. Notice that the forces in Eqs.~(\ref{er}) and (\ref{fr}) are the negative of those in Eqs.~(\ref{ei}) and (\ref{fi}), as they should be for the present case, since the combination of orbital and spin rotations of our vortex around its core leaves it invariant, producing no forces.

The total resistivity tensor (\ref{damping}) is (in the cylindrical coordinates) 
\ben
\hat{\gamma}=\gamma+\eta\left(d_{rr}+d_{\phi\phi}\right)+\eta^\prime\hat{d}+\frac{q}{\rho}\hat{b}=\left(\begin{array}{cc}\gamma_r & \gamma_\perp\\-\gamma_\perp & \gamma_\phi\end{array}\right)\,,
\een
where
\begin{align}
 \gamma_r&=\gamma+(\eta+\eta^\prime)(\p_r\theta)^2+\eta\left(\frac{\sin\theta}{r}\right)^2\,,\nn
\gamma_\phi&=\gamma+\eta(\p_r\theta)^2+(\eta+\eta^\prime)\left(\frac{\sin\theta}{r}\right)^2\,,\nn
\gamma_\perp&=-\frac{q}{\rho}\frac{\p_r\cos\theta}{r}\,.
\end{align}
Here, the two diagonal components, $\gamma_r$ and $\gamma_\phi$, describe the (dissipative) anisotropic resistivity, while the off-diagonal component, $\gamma_\perp$, captures what is called the topological Hall effect.\cite{brunoPRL04}

\begin{figure}[htbp]
\includegraphics [width=0.7\linewidth]{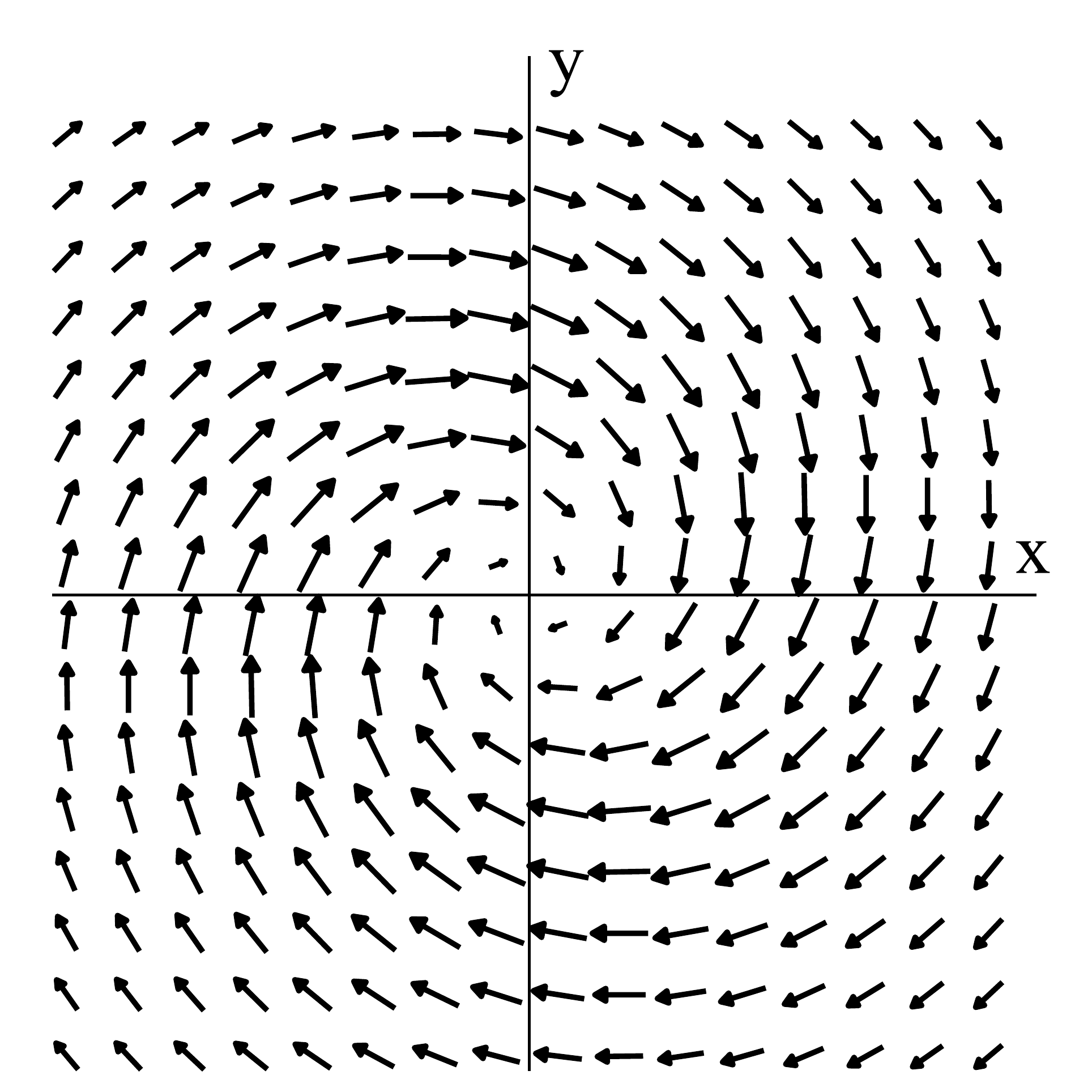}
\caption{We plot here the current in Eq.~(\ref{cc}) (all parameters set to 1). Near the core, the current spirals inward and charges build up at the center (which is allowed for our compressible fluid).}
\label{current}
\end{figure}

In the drift approximation, Eq.~(\ref{da}), the current-density field $\mbf{j}=j_r\mbf{\hat{r}}+j_\theta\bs{\hat{\theta}}$ is given by
\begin{align}
\mbf{j}&=\hat{\gamma}^{-1}q(\mbf{e}+\mbf{f}^{(\beta)})\,, \nn
\left(\begin{array}{cc} j_r\\ j_ \phi \end{array}\right)&=q\omega\hat{\gamma}^{-1}\left(\begin{array}{c}\p_r\cos\theta\\-\beta\sin^2\theta/r\end{array}\right)\nn
&\hspace{-0.5cm}=-\frac{q\omega\sin\theta}{\gamma_r\gamma_\phi+\gamma^2_\perp}\left(\begin{array}{cc}\gamma_\phi & -\gamma_\perp\\\gamma_\perp & \gamma_r\end{array}\right)\left(\begin{array}{c}\p_r\theta\\\beta\sin\theta/r\end{array}\right)\,.
\label{cc}
 \end{align}
More explicitly, we may consider a profile $\theta=\pi(1-e^{-r/a})/2$, where $a$ is the radius of the vortex core. The corresponding current (\ref{cc}) is sketched in Fig.~\ref{current}.

We note that the fictitious magnetic field $\epsilon^{ijk}b_{jk}/2$ points everywhere in the $z$ direction, its total flux through the $xy$ plane being given by 
\ben
\mcal{F}=\int d\phi dr(rb_{r\phi})=-\int d\phi dr (\p_\phi\varphi\p_r\cos\theta)=2\pi\,.
\een
Note that the integrand is just the Jacobian of the map from the plane to the sphere defined by the spin-texture field:
\ben
(\theta (\mbf{r}),\varphi(\mbf{r})):R^2\to S^2\,.
\label{RS}
\een
This reflects the fact that the fictitious magnetic flux is generally a topological invariant, corresponding to the $\pi_2$ homotopy group of the mapping (\ref{RS}).\cite{belavinJETPL75,volovikJPC87}

\subsection{Anisotropic resistivity of a 3D spiral}

\begin{figure}[htbp]
\includegraphics [width=\linewidth]{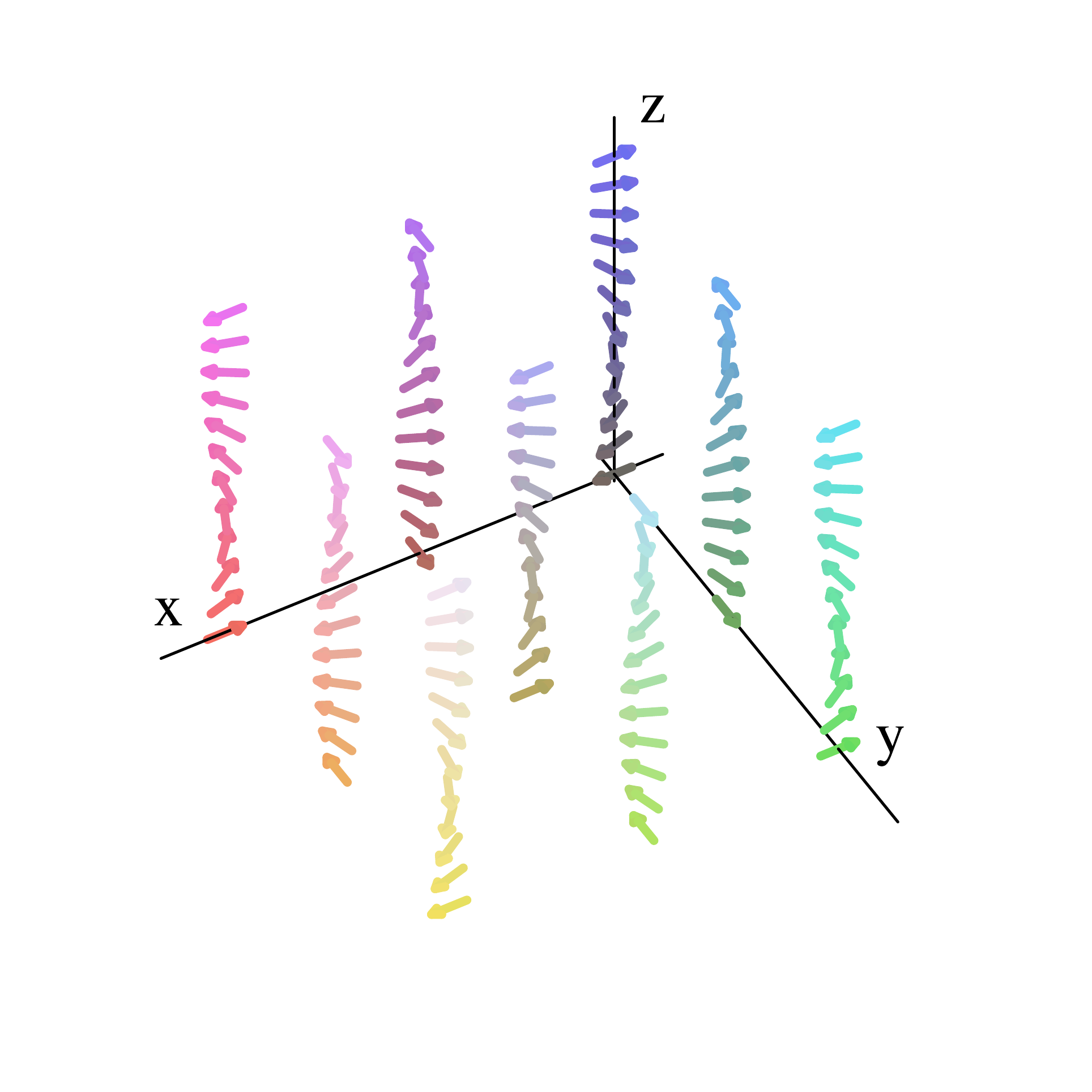}
\caption{(Color online) A set of spin spirals which is topologically trivial because $\boldsymbol{\nabla}\theta=0$ (and equivalent to the spin helix, Fig.~\ref{helix}, up to a global real-space rotation), hence the fictitious magnetic field $\mbf{b}$, Eq.~(\ref{tensors}), is zero. There is, however, an anisotropic texture-dependent resistivity with finite off-diagonal components, Eq.~(\ref{damping1}).}
\label{spiral}
\end{figure}

Consider the texture described by $\p_i\mbf{m}=\kappa_i\mbf{\hat{z}}\times\mbf{m}$, where the spatial rotation stays in the $xy$ plane, but the wave vector $\bs{\kappa}$ can be in any direction. The spin texture forms a transverse helix in the $z$ direction and a planar spiral in the $x$ and $y$ directions. Fig.~\ref{spiral} shows such a configuration for $\bs{\kappa}$ pointing along $(\mbf{x}+\mbf{y}+\mbf{z})/\sqrt{3}$. The fictitious magnetic field $\mbf{b}$ vanishes, but the anisotropic resistivity still depends nontrivially on the spin texture:
\begin{align}
\gamma_{ij}&=\left[\gamma+\eta(\p_k\mbf{m})^2\right]\de_{ij}+\eta^\prime\p_i\mbf{m}\cdot\p_j\mbf{m}\nn
&= (\gamma+\eta\kappa^2)\de_{ij}+\eta^\prime\kappa_i\kappa_j\,,
\label{damping1}
\end{align}
which, according to $\mbf{j}=\hat{\gamma}^{-1}\mbf{E}$, would give a transverse current signal for an electric field applied along the Cartesian axes $x$, $y$, or $z$.

\section{summary}
\label{SUM}

We have developed semi-phenomenologically the hydrodynamics of spin and charge currents interacting with collective magnetization in metallic ferromagnets, generalizing the results of Ref.~\onlinecite{tserkovPRB09md} to three dimensions and compressible flows.  Our theory reproduces known results such as the spin-motive force generated by magnetization dynamics and the dissipative spin torque, albeit from a different viewpoint than previous microscopic approaches. Among the several new effects predicted, we find both an isotropic and an anisotropic texture-dependent resistivity, Eq.~(\ref{damping}), whose contribution to the classical (topological) Hall effect should be described on par with that of the fictitious magnetic field. By calculating the dissipation power, we give a lower bound on the spin-texture resistivity as required by the second law of thermodynamics. We find a more general form, including a term of order $\beta$, of the texture-dependent correction to nonlocal Gilbert damping, predicted in Ref.~\onlinecite{tserkovPRB09md}. See Eq.~(\ref{Gtensor}).

Our general theory is contained in the stochastic hydrodynamic equations, Eqs.~(\ref{stochastic}), which we treated in the highly compressible limit. The most general situation is no doubt at least as rich and complicated as the classical magnetohydrodynamics. A natural extension of this work is the inclusion of heat flows and related thermoelectric effects, which we plan to investigate in a future work.  Although we mainly focused on the halfmetallic limit in this paper, our theory is in principle a two-component fluid model and allows for the inclusion of a fully dynamical treatment of spin densities and associated flows.\cite{nTwo} Finally, our hydrodynamic equations become amenable to analytic treatments when applied to the important problem of spin-current driven dynamics of magnetic solitons, topologically stable objects that can be described by a small number of collective coordinates, which we will also investigate in future work. 

\acknowledgments

We are grateful to Gerrit~E.~W. Bauer, Arne Brataas, Alexey~A. Kovalev, and Mathieu Taillefumier for stimulating discussions. This work was supported in part by the Alfred~P. Sloan Foundation and the NSF under Grant No. DMR-0840965.

\appendix

\section{Many-body action}
\label{mba}

We can formally start with a many-body action, with Stoner instability built in due to short-range repulsion between electrons:\cite{duinePRB07fk}
\begin{align}
S[\bar{\psi}_\s(\mbf{r},t),\psi_\s(\mbf{r},t)]=\int_\mathcal{C}dt\int d^3r&\nn
&\hspace{-4.5cm}\left[\hat{\psi}^+\left(i\hbar\partial_t+{\hbar^2\over2m_e}\del^2\right)\hat{\psi}-U\bar{\psi}_\uparrow\bar{\psi}_\downarrow\psi_\downarrow\psi_\uparrow\right]\,,
\end{align}
where time $t$ runs along the Keldysh contour from $-\infty$ to $\infty$ and back. $\bar{\psi}_\s$ and $\psi_\s$ are mutually independent Grassmann variables parametrizing fermionic coherent states and $\hat{\psi}^+=(\bar{\psi}_\uparrow,\bar{\psi}_\downarrow)$ and $\hat{\psi}=(\psi_\uparrow,\psi_\downarrow)^T$. The four-fermion interaction contribution to the action can be decoupled via Hubbard-Stratonovich transformation, after introducing auxiliary bosonic fields $\phi$ and $\bs{\Delta}$:
\begin{align}
e^{iS_U/\hbar}&=\exp\left(-\frac{i}{\hbar}\int_\mathcal{C}dt\int d^3rU\bar{\psi}_\uparrow\bar{\psi}_\downarrow\psi_\downarrow\psi_\uparrow\right)\nn
&=\int D[\phi(\mbf{r},t),\bs{\Delta}(\mbf{r},t)]\exp\left({i\over\hbar}\int_\mathcal{C}dt\int d^3r\right.\nn
&\hspace{0.6cm}\left.\left[{\phi^2\over 4U}-{\Delta^2\over 4U}-{\phi\over2}\hat{\psi}^+\hat{\psi}+{\bs{\Delta}\over2}\hat{\psi}^+\hat{\bs{\s}}\hat{\psi}\right]\right)\,.
\label{SU}
\end{align}
In obtaining this result, we decomposed the interaction into charge- and spin-density pieces:
\ben
\bar{\psi}_\uparrow\bar{\psi}_\downarrow\psi_\downarrow\psi_\uparrow={1\over4}(\hat{\psi}^+\hat{\psi})^2-{1\over4}(\hat{\psi}^+\mbf{m}\cdot\hat{\bs{\s}}\hat{\psi})^2\,,
\een
where $\mbf{m}$ is an arbitrary unit vector. It is easy to show that $\langle\phi(\mbf{r},t)\rangle=U\langle\hat{\psi}^+(\mbf{r},t)\hat{\psi}(\mbf{r},t)\rangle$ and $\langle\bs{\Delta}(\mbf{r},t)\rangle=U\langle\hat{\psi}^+(\mbf{r},t)\hat{\bs{\s}}\hat{\psi}(\mbf{r},t)\rangle$, when properly averaging over the coupled quasiparticle and bosonic fields.

The next step in developing mean-field theory is to treat the Hartree potential $\phi(\mbf{r},t)$ and Stoner exchange $\bs{\Delta}(\mbf{r},t)\equiv\Delta(\mbf{r},t)\mbf{m}(\mbf{r},t)$ fields in the saddle-point approximation. Namely, the effective bosonic action
\ben
S_{\rm eff}[\phi(\mbf{r},t),\bs{\Delta}(\mbf{r},t)]=-i\hbar\ln\int D[\hat{\psi}^+,\hat{\psi}]e^{\frac{i}{\hbar}S(\hat{\psi}^+,\hat{\psi};\phi,\bs{\Delta})}
\label{Seff}
\een
is minimized, $\delta S_{\rm eff}=0$, in order to find the equations of motion for the fields $\phi$ and $\bs{\Delta}$. In the limit of sufficiently low electron compressibility and spin susceptibility, the charge- and spin-density fluctuations are suppressed, defining mean-field parameters $\bar{\phi}$ and $\bar{\Delta}$. Since a constant $\bar{\phi}$ only shifts the overall electrochemical potential, it is physically inconsequential. Our theory is designed to focus on the remaining soft (Goldstone) modes associated with the spin-density director $\mathbf{m}(\mbf{r},t)$, while $\phi(\mbf{r},t)$ and $\Delta(\mbf{r},t)$ are in general allowed to fluctuate close to their mean-field values $\bar{\phi}$ and $\bar{\Delta}$, respectively. The saddle-point equation of motion for the collective spin direction $\mathbf{m}(\mbf{r},t)$ follows from $\delta_\mbf{m}S_{\rm eff}[\mbf{m}]=0$, after integrating out electronic degrees of freedom. Because of the noncommutative matrix structure of the action (\ref{SU}), it is still a nontrivial problem. The problem simplifies considerably in the limit of large exchange splitting $\Delta$, where we can project spins on the local magnetic direction $\mbf{m}$. This lays the ground to the formulation discussed in Sec.~\ref{QA}, where the collective spin-density field parametrized by the director $\mbf{m}(\mbf{r},t)$ interacts with the spin-up/down free-electron field. The resulting equations of motion constitute the self-consistent dynamic Stoner theory of itinerant ferromagnetism.

In the remainder of this appendix, we explicitly show that the semiclassical formalism developed in Secs.~\ref{QA}-\ref{CEM} is equivalent to a proper field-theoretical treatment. The equation of motion for the spin texture follows from extremizing the effective action with respect to variations in $\mbf{m}$.  Because of the constraint on the magnitude of $\mbf{m}$, its variation can be expressed as $\delta\mbf{m}=\delta \boldsymbol{\theta}\times\mbf{m}$, with $\delta\boldsymbol{\theta}$ being an arbitrary infinitesimal vector, so that the equation of motion is given by $\mbf{m}\times\delta_\mbf{m}S_{\rm eff}=0$: 
\begin{align}
0&=\mbf{m}\times\delta_\mbf{m}S_{\rm eff}\nn
&={1\over\mathcal{Z}}\int D[\hat{\psi}^+,\hat{\psi}]\left(\mbf{m}\times\delta_\mbf{m}S\right)e^{\frac{i}{\hbar}S[\hat{\psi}^+,\hat{\psi};\phi,\bs{\Delta}]}\nn
&=\sum_{\s\mu}(\mbf{m}\times\de_\mbf{m}a_{\s\mu})\left\langle\frac{\partial S}{\partial a_{\s\mu}}\right\rangle-\mbf{m}\times\delta_\mbf{m}F\,,
\label{EOM}
\end{align}
where $\mathcal{Z}=\int D[\hat{\psi}^+,\hat{\psi}]e^{\frac{i}{\hbar}S[\hat{\psi}^+,\hat{\psi};\phi,\bs{\Delta}]}$ and we have used the path-integral representation of the vacuum expectation value. $a_{\s\mu}$ are the spin-dependent gauge potentials (\ref{aa}) and $F$ the spin exchange energy, appearing after we project spin dynamics on the collective field $\bs{\Delta}$. Equation~\eqref{EOM} may be expressed in terms of the hydrodynamic variables of the electrons. Defining spin-dependent charge and current densities, $j_\s^\mu=(\rho_\s,\mbf{j}_\s)$,  by 
\begin{align}
\rho_\s&=\left\langle\partial S\over\p a_\s\right\rangle= \langle\bar{\psi}_\s\psi_\s\rangle\,,\nn
\mbf{j}_\s&=\left\langle{\partial S\over\p\mbf{a}_\s}\right\rangle= {1\over m_e}{\rm Re}\left\langle\bar{\psi}_\s(-i\hbar\bs{\del}- \mbf{a}_\s)\psi_\s\right\rangle=\rho_\s\mbf{v}_\s\,,
\end{align}
Eq.~\eqref{EOM} reduces to the Landau-Lifshitz Eq.~\eqref{LL}. Minimizing action (\ref{Seff}) with respect to the $\phi$ and $\Delta$ fields gives the anticipated self-consistency relations:
\begin{align}
\phi(\mbf{r},t)&=U\langle\hat{\psi}^+(\mbf{r},t)\hat{\psi}(\mbf{r},t)\rangle=U(\rho_++\rho_-)\,,\nonumber\\
\Delta(\mbf{r},t)&=U\langle\hat{\psi}^+(\mbf{r},t)\hat{\s}_z\hat{\psi}(\mbf{r},t)\rangle=U(\rho_+-\rho_-)\,.
\end{align}

\section{The monopole gauge field}
\label{mgf}

Let $(\theta,\varphi)$ be the spherical angles of $\mbf{m}$, the direction of the local spin density, and $\hat{\chi}_\s$ be the spin up/down ($\s=\pm$) spinors given by, up to a phase,
 \begin{align} 
  \hat{\chi}_+(\theta,\varphi) &= \left( \begin{array}{c}   
   \cos{\theta\over2} \\  e^{i\varphi}\sin{\theta\over2}\end{array}\right)\,,\nn
   \hat{\chi}_-(\theta,\varphi)&=\hat{\chi}_+(\pi-\theta, \varphi+\pi)=\left( \begin{array}{c}   \sin{\theta\over2}  \\   -e^ {i\varphi} \cos {\theta\over2} \end{array} \right )\,.
   \label{spinors}
\end{align}
The spinors are related to the spin-rotation matrix $\hat{\mathcal{U}}(\mbf{m})$ by $\hat{\chi}_\sigma=\hat{\mathcal{U}}|\sigma\rangle$. The gauge field in $\mbf{m}$ space, which enters Eq.~(\ref{gauge fields}), is thus given by
\ben
\mbf{a}^{\rm mon}_\s(\theta,\varphi)=-i\hbar\hat{\chi}^\dag_\s\p_\mbf{m}\hat{\chi}_\s={\hbar\over2}\left(\frac{1-\s\cos\theta}{\sin\theta}\right){\bs{\hat{\varphi}}}\,,
\label{Amon}
\een
where we used the gradient on a unit sphere: $\p_\mbf{m}=\bs{\hat{\theta}}\partial_\theta+\bs{\hat{\varphi}}\partial_\varphi/\sin\theta$.  
The magnetic field corresponding to this vector potential [extended to three dimensions by $\mbf{a}(\mathbf{m})\to\mbf{a}(\theta,\varphi)/m$] is given on the unit sphere by
\ben
\p_\mbf{m}\times\mbf{a}_\s^{\rm mon}=\p_\mbf{m}\times(a_\varphi\bs{\hat{\varphi}})={\mbf{m}\over\sin\theta}\partial_\theta(\sin\theta a_\varphi)={\s\hbar\over2}\mbf{m}\,.
\label{monopole}
\een
It follows from Eqs.~\eqref{gauge fields} and \eqref{Amon} that the spin-dependent real-space gauge fields are given by
\ben
a_{\s\mu}=-\frac{\hbar}{2}\partial_\mu\varphi (1-\s\cos\theta)\,.
\label{AA}
\een
Notice that the $\s=\pm$ monopole field (\ref{Amon}), as well as the above gauge fields, are singular on the south/north pole (corresponding to the Dirac string). This is what allows a magnetic field with finite divergence.  Any other choice of the monopole gauge field (\ref{Amon}) would correspond to a different choice of the spinors (\ref{spinors}), translating into a gauge transformation of the fields (\ref{AA}). This is immediately seeing by noticing that $\mbf{a}_\s^{\rm mon}(\mbf{m})\to\mbf{a}_\s^{\rm mon}(\mbf{m})+\p_\mbf{m}f_\s(\mbf{m})$ corresponds to $a_{\s\mu}(\mbf{r},t)\to a_{\s\mu}(\mbf{r},t)+\partial_\mu f_\s(\mbf{m}(\mbf{r},t))$.


\begin{thebibliography}{58}
\bibitem{nExchange} Coupling of this exchange energy to the electronic density fluctuations can also be treated systematically. However, since it does not lead to any significant effects, while at the same time unnecessarily complicating our discussion, we will disregard it in the following.
\bibitem{nEnergy} Generally, the phenomenological free energy of a ferromagnet is given by $F[\mbf{m}]=A(\p_i\mbf{m})^2/2 +U_{\rm ani}(\mbf{m})-M\mbf{h}_{\rm ext}\cdot\mbf{m}$, where $A$ is the effective stiffness, $U_{\rm ani}(\mbf{m})=U_{\rm ani}(-\mbf{m})$ is an anisotropic potential due to magnetostatic and crystalline fields, $M$ is the equilibrium magnetization, and $\mbf{h}_{\rm ext}$ is an external magnetic field.
\bibitem{nTwo} In the more general two-band model of spin-up and spin-down electrons, we would have to develop a two-fluid effective theory, with spin-flip scattering between the two fluid components. The phenomenology simplifies, however, reducing formally to the halfmetallic case, in the long-wavelength low-frequency limit: $\omega\tau_{\rm sf}\ll1$ and $k\lambda_{\rm sd}\ll1$, where $\omega$, $k$ are the characteristic frequency, wave number of the magnetohydrodynamics and $\tau_{\rm sf}$, $\lambda_{\rm sd}\propto\sqrt{\tau_{\rm sf}}$ are the spin-flip time, spin-diffusion length. In this limit, it may be possible to describe the hydrodynamic state of the system by the spin-texture field, the charge-density distribution, and the charge-current field. If any out-of-equilibrium spin imbalance decays sufficiently fast, therefore, we only need to retain a one-fluid description for the charge flows. The key phenomenological modification is then to introduce a material-dependent dimensionless ``spin-polarization" parameter $p$, such that $q\to pq$ in the following equations of motion. Namely, the effective charge that couples the electronic particle-number flux densities $\mbf{j}$ with the spin-texture gauge field is renormalized by $p$. While in the halfmetallic limit $p=1$ and in normal metals $p=0$, we may expect some intermediate value in realistic multiple-band ferromagnets with fast spin relaxation.
\bibitem{nCC} To make contact with Ref.~\onlinecite{tserkovPRB09md}, define the canonical current as $\mcal{J}= \oint_C \mbf{p}\cdot  d\mbf{l}$, for an arbitrary closed curve $C$. Its equation of motion is given by $\delt\mcal{J}=q\oint_Cd\mbf{l}\cdot(\mbf{v}\times\mbf{b})$. If the curve $C$ coincides with a quasi-1D wire, then $\mbf{v}\| d\mbf{l}$ and we recover the reactive equation of Ref.~\onlinecite{tserkovPRB09md}: $\delt\mcal{J}=0$.
\bibitem{nEM} Similarly to $q$ that should generally be viewed as a material-dependent phenomenological parameters, the electron mass $m_e$ from now on is also an effective parameter, which is not necessarily identical with the free-electron mass.
\bibitem{nViscosity} We are not including the ordinary hydrodynamic viscosity in our treatment.
\end{thebibliography}
\end{document}